# MEMS MAGNETOMETER USING MAGNETIC FLUX CONCENTRATORS AND PERMANENT MAGNETS

*Federico Maspero [1,\*], Gabriele Gatani,[3] Simone Cuccurullo[2] and Riccardo Bertacco[1,2]*
[1]CNR-Istituto di Fotonica e Nanotecnologie, Milano, Italy,
[2]Politecnico di Milano, Milano, Italy and
[3]Politecnico di Torino, Torino, Italy

## ABSTRACT
A novel frequency modulated MEMS magnetometer is presented. The new device is inspired by the magnetic force microscopy technique, it combines standard MEMS technology with magnetic elements to obtain a magnetometer with potential for low power consumption and high performance. The analytical model and the characterization of the first proof-concept device are presented.

## INTRODUCTION
Commercial integrated magnetometers rely on Magneto Resistive technology (AMR, GMR, TMR). These devices are combined with MEMS accelerometers and gyroscope in the same package to form Inertial Measurement Units (IMU). The possibility of integrating magnetometers on the same silicon die of the inertial sensors is one of the reasons that pushed the development of MEMS magnetometers in the last years. Several types of MEMS magnetic sensors were proposed in the past [1]–[9] and they can mainly be divided in two groups: (i) devices which incorporate magnetic materials in order to produce a field-dependent mechanical effect [1]–[6] (ii) and Lorentz force magnetometers [7]–[9]. The latter do not need magnetic material and are usually compatible with existing MEMS processes, therefore having an advantage over the first group of sensors. On the other hand, Lorentz force magnetometers suffer from a tradeoff between length of the current path (device size), current consumption and sensitivity.

In this work we propose a novel type of MEMS magnetometer combining permanent magnets and magnetic field concentrators (MFCs). As shown in the following sections, this type of magnetometers can potentially achieve noise level in the nT/√Hz while having current consumption of a MEMS resonator which is now in the order of few µA [10].

## THE DEVICE
**Magnetic force microscopy**
The MEMS magnetometer proposed in this work is inspired by the concept underlying magnetic force microscopy (MFM) (Fig.1).

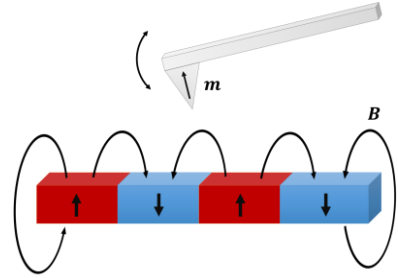

*Figure 1 - Schematic representation of the MFM working principle.*

MFM is a technique used to map the magnetic stray field produced by a magnetized sample. Such mapping is a useful tool to reconstruct the magnetization of the sample down to single magnetic domains.

As shown in Fig.1, the MFM technique uses a scanning probe with a magnetized tip, kept in oscillation close to the substrate. A tip magnetized along x interacts with the stray field according to the following equation:

$$F_s = m_x \frac{\partial B_x}{\partial s} \quad (1)$$

Where $F_s$ is the s-component of the force acting on the magnetized element, $m_x$ is the magnetic dipole moment of the tip, $B_x$ is the x-component of the stray magnetic field and $s$ is the displacement direction of the cantilever.

The non-uniformity of the stray field makes the force position-dependent, thus leading to an equivalent magnetic stiffness ($k_{mag}$):

$$k_{mag} = \frac{\partial F}{\partial s} = m_x \frac{\partial^2 B_x}{\partial s^2} \quad (2)$$

The cantilever is affected by the equivalent stiffness and changes its resonance frequency according to:

$$\Delta \omega_R \approx m_x \frac{\omega_R}{2k} \frac{\partial^2 B_x}{\partial s^2} \quad (3)$$

where $\Delta\omega_R$ is the resonance frequency shift, $\omega_R$ is the resonance frequency expressed in rad/s and $k$ the elastic constant of the structure, assumed to be much greater than $k_{mag}$.

The frequency shift is used to determine the direction of the field gradient and eventually its amplitude. Depending on the relative orientation between $m_x$ and the stray magnetic field, this phenomenon can lead to both softening and hardening of the device stiffness, this enables to discern between positive and negative field gradients.

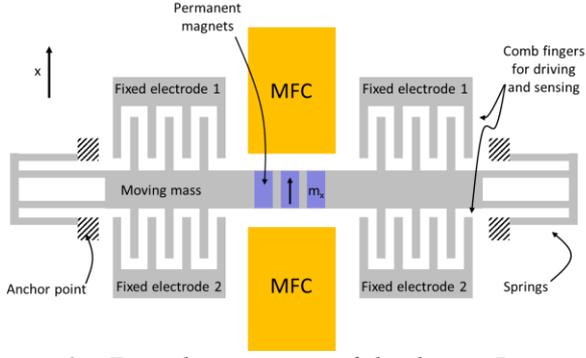

*Figure 2 - Top schematic view of the device. Permanent magnets are patterned on top of a MEMS resonator. A field gradient is produced close to the device by a pair of MFCs.*

**MEMS magnetometer**

The same concept is exploited in the proposed sensor (Fig.2), a MEMS resonator is functionalized with a magnetic material patterned on top of the moving element in order to sense the magnetic field gradient. As standard magnetometers must sense fields, not gradients, a conversion of an external field into an appropriate gradient with high second derivative is needed.

For this purpose, the magnetic materials on top of the resonator is placed in the gap between two magnetic flux concentrators [11]. The MFCs amplify the incoming magnetic field and produce a non-uniform field in the constriction, whose second derivative is directly related to the external field.

**Device sensitivity and resolution**

To estimate the sensitivity and resolution of the device, the second derivative of the field between the MFCs is needed.

Referring to Fig.2, let us consider an external field $B_a$ applied along x, i.e. along the edge of rectangular MFCs made of thin films; the magnetization of the MFC along x is given by [12]:

$$M = \frac{\mu_r - 1}{1 + N_x(\mu_r - 1)} \frac{B_a}{\mu_0}, \quad (4)$$

where $\mu_0$ is the vacuum permeability, $\mu_r$ is the relative permeability of the magnetic material and $N_x$ is its demagnetizing factor [13].

Once defined the magnetization of the MFCs, it is possible to analytically obtain the approximated value of the magnetic field second derivative at the center of the gap [13]:

$$\frac{\partial^2 B_x}{\partial x^2} \approx \frac{8}{\pi} \frac{\mu_r - 1}{1 + N_x(\mu_r - 1)} \frac{t}{g^3} B_a \equiv \gamma B_a, \quad (5)$$

where $t$ is the thickness of the two rectangular MFCs and $g$ is the gap between them. The relation between the applied field and the second derivative $\frac{\partial^2 B_x}{\partial x^2}$ is linear via the coefficient $\gamma$ which depends on the magnetic and geometric properties of the MFCs.

The field sensitivity of the device is given by:

$$\frac{\Delta \omega_R}{B_a} = m_x \gamma \frac{\omega_R}{2k}, \quad (6)$$

Equivalently, the sensitivity can be expressed in terms of phase shift per unit field [14]:

$$S_\phi = \frac{\Delta \phi}{B_a} = \frac{2Q}{\omega_R} \frac{\Delta \omega_R}{B_a} = m_x \gamma \frac{Q}{k}, \quad (7)$$

where $Q$ is the quality factor of the resonator.

The expected field equivalent noise of the device can now be evaluated by taking the ratio between the phase noise and the sensitivity:

$$N_B = \frac{N_\phi}{S_\phi}. \quad (8)$$

If the noise introduced by the detection system is neglected, $N_\phi$ is given only by the thermomechanical noise of the resonator. For a noise bandwidth much smaller than $\frac{\omega_R}{Q}$, the phase equivalent noise can be approximated with [14]:

$$N_\phi \approx \sqrt{\frac{4k_B T}{\bar{x}^2 k} \frac{Q}{\omega_R}} \quad (9)$$

where $k_B$ is the Boltzmann constant, $T$ is the absolute temperature and $\bar{x}$ is the average displacement of the resonator.

Finally, the estimated field equivalent noise is:

$$N_B = \frac{N_\phi}{S_\phi} = \frac{\sqrt{4k_B T b}}{\bar{x}} \frac{1}{m_x \gamma} \quad (10)$$

where $b$ is the damping coefficient.

**Fabrication**

The fabrication of first prototypes of the magnetic sensor was carried out at PoliFAB, the cleanroom of Politecnico di Milano.

The devices are fabricated starting from a silicon on insulator substrate with 5um of doped single crystal silicon deposited on top of the 1um of sacrificial oxide (1).

Three layers are subsequently deposited and patterned using lift-off technique: the metal layer for the electric contacts, the hard-magnetic material for the permanent

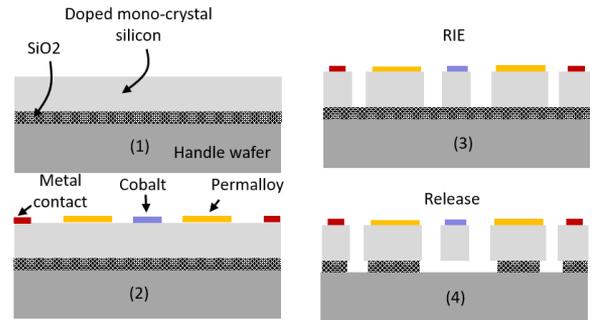

*Figure 3 – Schematic view of the main fabrication steps.*

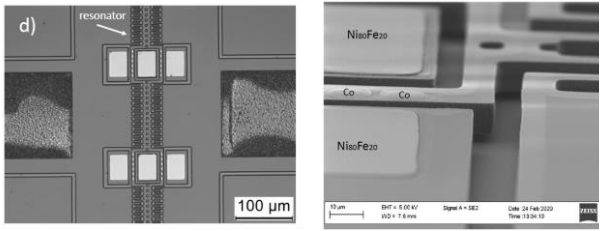

*Figure 4 (Left) Optical top view of a device; (right) Image of the magnetic elements using a scanning electrons microscope.*

magnet (Cobalt) and the soft-magnetic material for the MFCs (permalloy $Ni_{80}Fe_{20}$) (2).

The substrate is etched using Reactive Ion Etching to define the MEMS structure (3).

Finally, the sensor is released in buffered oxide etch solution (4).

## CHARACTERIZATION
### Setup

The preliminary tests on the device have been performed at ambient pressure using a manual probe station while applying an external magnetic field through an electromagnet. The device has three electrical connection: two to the fixed electrode on both sides of the resonator and one to the moving mass (Fig.1).

The device is actuated at its resonance frequency and the phase is monitored as a function of the applied field.

The actuation is performed by applying a sinusoidal voltage on one of the fixed electrodes at half of the device resonance frequency ($\omega_{AC} = \omega_R/2$) while the other fixed electrode is polarized with a DC voltage. The signal produced by the displacement of the device at $\omega_R$, due to the change of the capacitance between fixed and movable parts, is collected by a transimpedance amplifier (Zurich Instrument HF2TA) connected to the moving mass electrode. The power supply of the electromagnet is driven with a low frequency sinusoidal signal (0.25Hz), thus sweeping the applied filed between negative and positive values.

AC signal generation and demodulation is performed with the Lock-in amplifier Zurich Instrument HF2LI, while the DC voltage is generated with an external power supply.

### Preliminary measurement

Fig.5a shows the reference applied field and phase shift versus time. As a field is applied the device resonance frequency changes leading to a phase shift.

Fig.5b shows the phase shift versus applied field for different amplitudes of magnetic excitation. The graph is obtained by averaging over several cycles and bandpass filtering the frequency of interest. From this graph two aspects are visible: (i) the incoming field is shaped by the MFC transfer function leading to a hysteresis in the demagnetizing field i.e. phase shift; (ii) the magnetization of the permanent magnets increases or decreases with the applied field, leading to an increase or decrease of phase shift for large fields.

The obtained phase shift per unit of field is computed by linearly fitting the central part of the hysteresis cycle. The sensitivity measured on this specific device is $S_{\phi_m} = 0.51 \pm 0.03$ °/mT. The theoretical sensitivity is calculated using equation (5)(7) and the values reported in Tab.1: $S_{\phi_{th}} = 0.19$ °/mT. The measured and theoretical sensitivity are in the same order of magnitude. Two aspects should be pointed out: (i) the analytical model overestimated the second derivative assuming a uniform magnetization of the material, which is not realistic; (ii) the DC-voltage used to sense the device induces a position offset moving the suspended mass closer to one of the MFC and increasing the value of the magnetic force. Both these aspects lead to an increase of the theoretical and measured sensitivity respectively. Nevertheless, the analytical model presented in this work offers a useful tool to predict the ultimate device sensitivity and resolution.

The latter is not experimentally investigated in this work as the device should be tested in vacuum condition. However, a theoretical estimated can be given based on the preliminary measurement and on equation (10), assuming typical values of quality factor (Q=10000) and displacement ($\bar{x} = 1$ $\mu$m) of this type of MEMS resonator [8]. The expected noise floor for this type of sensor is equal to $N_{Bth} = 13$ nT/√Hz. This value could be largely improved for instance by increasing the thickness of the magnetic material deposited on top of the device, which is now only 100nm.

Finally, (not shown in the graphs) it was observed that by switching the magnetization the phase shift for a given field is reversed. This is in line with the theoretical prediction, as the force depends on the orientation between the magnetic dipole moment and the field gradient.

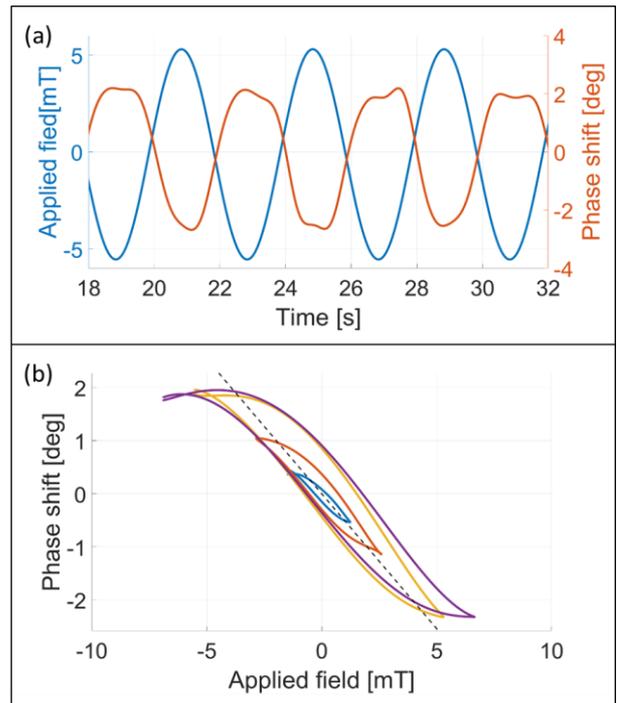

*Figure 5 - (a) Plot of device phase shift and reference actuation voltage versus time; (b) Phase shift versus applied field for different amplitude of sinusoidal excitation. The black dashed line represents the average slope of the linear part of the hysteresis curves.*

Table 1- Measured and designed values of the characterized MEMS magnetometer

| | |
|---|---|
| $\mu_r$ | 500 |
| $N_x$ | $2.6 \cdot 10^{-3}$ |
| $t$ | 100 nm |
| $g$ | 20 $\mu$m |
| $\gamma$ | $6.9 \cdot 10^9$ m$^{-2}$ |
| $m_x$ | $5.34 \cdot 10^{-11}$ Am$^2$ |
| $\omega_R$ | 12.4 kHz |
| $k$ | 1.1 N/m |
| $Q$ | 9.8 |

## CONCLUSIONS

A novel MEMS magnetometer was presented. The new sensor is inspired to MFM, a technique where the frequency of a magnetized scanning probe is changed by the magnetic interaction with an external field. By integrating magnetic elements such as MFCs and permanent magnets into a basic MEMS process we fabricated the first proof of concept device. The analytical model of the sensor was given together with the first experimental results. The device showed a response to the magnetic field in line with theoretical predictions and it is expected to reach noise floor in the order of few nT/√Hz.

## ACKNOWLEDGEMENTS

This work was carried out in the frame of the European FET-Open Project OXiNEMS. This project has received funding from the European Union's Horizon 2020 research and innovation programme under grant agreement No 828784.

## REFERENCES

[1] D. DiLella *et al.*, "A micromachined magnetic-field sensor based on an electron tunneling displacement transducer," *Sensors and Actuators A: Physical*, vol. 86, no. 1, pp. 8–20, Oct. 2000, doi: 10.1016/S0924-4247(00)00303-4.

[2] R. Levy, T. Perrier, P. Kayser, B. Bourgeteau, and J. Moulin, "A micro-resonator based magnetometer," *Microsyst Technol*, vol. 23, no. 9, pp. 3937–3943, Sep. 2017, doi: 10.1007/s00542-015-2790-2.

[3] S. Brugger, P. Simon, and O. Paul, "Field Concentrator Based Resonant Magnetic Sensor," in *2006 IEEE SENSORS*, Oct. 2006, pp. 1016–1019, doi: 10.1109/ICSENS.2007.355797.

[4] Y. Hui, T. Nan, N. X. Sun, and M. Rinaldi, "High Resolution Magnetometer Based on a High Frequency Magnetoelectric MEMS-CMOS Oscillator," *Journal of Microelectromechanical Systems*, vol. 24, no. 1, pp. 134–143, Feb. 2015, doi: 10.1109/JMEMS.2014.2322012.

[5] J. Zhai, Z. Xing, S. Dong, J. Li, and D. Viehland, "Detection of pico-Tesla magnetic fields using magneto-electric sensors at room temperature," *Appl. Phys. Lett.*, vol. 88, no. 6, p. 062510, Feb. 2006, doi: 10.1063/1.2172706.

[6] B. Gojdka *et al.*, "Fully integrable magnetic field sensor based on delta-E effect," *Appl. Phys. Lett.*, vol. 99, no. 22, p. 223502, Nov. 2011, doi: 10.1063/1.3664135.

[7] M. Li, S. Sonmezoglu, and D. A. Horsley, "Extended Bandwidth Lorentz Force Magnetometer Based on Quadrature Frequency Modulation," *Journal of Microelectromechanical Systems*, vol. 24, no. 2, pp. 333–342, Apr. 2015, doi: 10.1109/JMEMS.2014.2330055.

[8] G. Laghi, C. R. Marra, P. Minotti, A. Tocchio, and G. Langfelder, "A 3-D Micromechanical Multi-Loop Magnetometer Driven Off-Resonance by an On-Chip Resonator," *Journal of Microelectromechanical Systems*, vol. 25, no. 4, pp. 637–651, Aug. 2016, doi: 10.1109/JMEMS.2016.2563180.

[9] R. Sunier, T. Vancura, Y. Li, K.- Kirstein, H. Baltes, and O. Brand, "Resonant Magnetic Field Sensor With Frequency Output," *Journal of Microelectromechanical Systems*, vol. 15, no. 5, pp. 1098–1107, Oct. 2006, doi: 10.1109/JMEMS.2006.880212.

[10] P. M. Drljača, F. Vincent, P.-A. Besse, and R. S. Popović, "Design of planar magnetic concentrators for high sensitivity Hall devices," *Sensors and Actuators A: Physical*, vol. 97–98, pp. 10–14, Apr. 2002, doi: 10.1016/S0924-4247(01)00866-4.

[11] X. Sun, L. Jiang, and P. W. T. Pong, "Magnetic flux concentration at micrometer scale," *Microelectronic Engineering*, vol. 111, pp. 77–81, Nov. 2013, doi: 10.1016/j.mee.2013.01.063.

[12] J. A. Osborn, "Demagnetizing Factors of the General Ellipsoid," *Phys. Rev.*, vol. 67, no. 11–12, pp. 351–357, Jun. 1945, doi: 10.1103/PhysRev.67.351.

[13] Z. J. Yang, T. H. Johansen, H. Bratsberg, G. Helgesen, and A. T. Skjeltorp, "Potential and force between a magnet and a bulk Y1Ba2Cu3O7 superconductor studied by a mechanical pendulum," *Supercond. Sci. Technol.*, vol. 3, no. 12, pp. 591–597, Dec. 1990, doi: 10.1088/0953-2048/3/12/004.

[14] K. L. Ekinci, Y. T. Yang, and M. L. Roukes, "Ultimate limits to inertial mass sensing based upon nanoelectromechanical systems," *Journal of Applied Physics*, vol. 95, no. 5, pp. 2682–2689, Feb. 2004, doi: 10.1063/1.1642738.

## CONTACT

*F. Maspero federico.maspero@polimi.it